\documentclass[aps,preprint,amssymb,superscriptaddress]{revtex4-1}
\usepackage{floatrow}
\newfloatcommand{capbtabbox}{table}[\FBwidth]
\usepackage{graphicx}% Include figure files
\usepackage{dcolumn}% Align table columns on decimal point
\usepackage{bm}% bold math
\usepackage{natbib}
\usepackage{CJK}
\usepackage[utf8]{inputenc}
\usepackage[T1]{fontenc}
\usepackage{mathptmx}
\usepackage{gensymb}
\usepackage{graphicx}% Include figure files
\usepackage{epstopdf}
\usepackage{dcolumn}% Align table columns on decimal point
\usepackage{bm}% bold math
\usepackage{amsmath}
\usepackage{mathtools}
\usepackage{relsize}
\usepackage{multirow}
\usepackage{lineno}
\usepackage[colorlinks]{hyperref}
\begin{document}

\title{Optimizing Finite Structures to Suppress the Photonic Density of States}% Force line breaks with \\

\author{Prakash Mishra}
\affiliation{Department of Electrical Engineering and Computer Science, Syracuse University, Syracuse, NY 13210}
\author{Sukhad  Dnyanesh Joshi}
\affiliation{Department of Electrical Engineering and Computer Science, Syracuse University, Syracuse, NY 13210}
\author{Aditya Bahulikar}
\affiliation{Department of Electrical Engineering and Computer Science, Syracuse University, Syracuse, NY 13210}
\author{Quintin A. Hatzis}
\affiliation{Department of Electrical Engineering and Computer Science, Syracuse University, Syracuse, NY 13210}
\author{M. Cenk Gursoy}
\affiliation{Department of Electrical Engineering and Computer Science, Syracuse University, Syracuse, NY 13210}
\author{Rodrick Kuate Defo}
\email{rkuatede@syr.edu}
\affiliation{Department of Electrical Engineering and Computer Science, Syracuse University, Syracuse, NY 13210}

\date{\today}% It is always \today, today,

\begin{abstract}
We propose a topology-optimization framework for optimizing finite structures of arbitrary shape by combining density-based methods with level-set approaches. We first optimize regular polygonal structures to suppress the photonic density of states and find that the best performing polygon is consistent with a tiling of space with hexagonal unit cells. We next show that introducing cavities into hexagonal structures further suppresses the photonic density of states, particularly when the cavity is also hexagonal. Such a result would find application in the design of fiber-optic cables. We then describe an approach for optimizing arbitrary $x$-simple or $y$-simple designs that can recover finite supercells of a hexagonal unit cell. Our approach can therefore discover the symmetry of photonic-crystal primitive unit cells that significantly suppress the photonic density of states for a given set of material parameters within a single optimization.
\end{abstract}

\maketitle

There have been extensive efforts to determine structures that create large photonic band gaps, which are frequency ranges for which electromagnetic waves are forbidden from penetrating a structure, both in terms of the basis elements that are used to tile space and in terms of the topology of the underlying space that is tiled~\cite{sigmund_geometric_2008,qian_isogeometric_2011,liew_photonic_2011,jin_photonic_2001,miyazaki_photonic_2003,edagawa_photonic_2008,rockstuhl_suppression_2009,florescu_designer_2009,imagawa_photonic_2010,rechtsman_amorphous_2011,itin_geometric_2025,hart_a_2012,wang_structural_2018,nguyen_isogeometric_2015,molesky_inverse_2018,men_bandgap_2010,men_fabrication_2014,men_robust_2014,kim_automated_2023}. We have previously observed that the creation of a photonic bandgap is equivalent to the suppression of the photonic density of states (DOS) over a frequency window~\cite{bahulikar_structure_2025,joshi_structure_2025}. In doing so, we had leveraged the formalism of Liang and Johnson~\cite{liang_formulation_2013,strekha_suppressing_2024,chao_physical_2022,chao_maximum_2022} to exactly capture the suppression of the photonic DOS by suppressing the sum of DOS values evaluated at complex frequencies~\cite{bahulikar_structure_2025,joshi_structure_2025}. Given continued interest in two-dimensional (2D) systems~\cite{mattheakis_epsilon_2016,kuate_methods_2021,kuate_strain_2016,shiang_ab_2015}, we focus on finite 2D structures that suppress the photonic DOS in this work. In particular, we first show that the regular polygonal finite structure that best suppresses the photonic DOS is consistent with the tiling of the finite region of space with the primitive unit cell of the photonic crystal with the largest photonic band gap. This result necessarily holds in the limit where the finite structure has linear dimensions greater than the Bragg length~\cite{hasan_finite_2018,neve-oz_bragg_2004,bahulikar_structure_2025,joshi_structure_2025} of the photonic crystal with the largest gap. Given this result, we next develop a topology optimization approach that can discover the shape of $x$-simple or $y$-simple finite structures that maximally suppress the photonic DOS. In the Bragg-length limit alluded to above, such an approach discovers the symmetry of photonic-crystal primitive unit cells that significantly suppress the photonic DOS for a given set of material parameters without having to perform a search over all possible space groups. 

There have been a number of approaches for topology optimization of photonic crystals, which include level-set descriptions~\cite{kao_maximizing_2005,burger_inverse_2004,burger_a_2005,he_incorporating_2007} and density-based methods~\cite{dobson_maximizing_1999,cox_band_2000,shen_large_2002,halkjaer_maximizing_2006,sigmund_geometric_2008,sigmund_systematic_2003,jensen_systematic_2004,watanabe_broadband_2006,liang_formulation_2013,hammond_unifying_2025} (involving continuous relaxations of the permittivity $\epsilon \in [\epsilon_{\text{min}},\epsilon_{\text{max}}]$ throughout the design region). Density-based methods often require postprocessing to ensure binarization of designs for application to real materials~\cite{hammond_unifying_2025} (though we have shown that for optimization over a full Brillouin zone binarization occurs naturally~\cite{joshi_structure_2025}), while level-set descriptions suffer from the inability to naturally create new inner fronts (voids for example) dynamically during the topology optimization process~\cite{seo_isogeometric_2010}. We therefore propose to combine level-set descriptions with density-based methods to address the shortcomings of both approaches. We note that we are not combining level-set descriptions with density-based methods in the sense employed by Hammond \textit{et al.}~\cite{hammond_unifying_2025} and others~\cite{svanberg_density_2013,hagg_on_2018,zhou_minimum_2015,qian_topological_2013,lazarov_length_2016}, who sought to smoothly transition density-based methods into level-set descriptions over the course of optimizations. Rather, our approach simultaneously performs a density-based optimization and a level-set optimization. Explicitly, the shape of the design region is treated with a level-set description, while the material within the design region is treated with a density-based method. In this manner, designs that are not easily attained by local algorithms in density-based methods can be discovered more quickly. We apply ideas from isogeometric topology optimization in order to implement our approach~\cite{hughes_isogeometric_2005,seo_isogeometric_2010,seo_shape_2010,qian_isogeometric_2011}.

In our optimizations, we consider the bandwidth-averaged analytically continued photonic DOS objective $(\tilde{\text{DOS}}_N(\omega_0,\Delta\omega))$~\cite{bahulikar_structure_2025,joshi_structure_2025}, 
\begin{align}
&\tilde{\text{DOS}}_N(\omega_0,\Delta\omega) \\&= -\frac{6}{\pi}\text{Im}\left[\frac{\sum_{n = 0}^{N-1}\left(e^{i(\pi+2\pi n)/(2N)}\right)\sum_j\int\tilde{\mathbf{J}}_{j}^*(\mathbf{r})\cdot\tilde{\mathbf{E}}_{j}\left(\left(\omega_0-\frac{\Delta\omega}{2}e^{i(\pi+2\pi n)/(2N)}\right),\mathbf{r}\right)\text{d}\mathbf{r}}{\csc\left(\frac{\pi}{2N}\right)}\right],\label{eq:finalresidue}
\end{align}
where $\omega_0$ is some central frequency, $\Delta\omega$ is some bandwidth (equivalently, the band gap), and $N$ is some integer (equal to the number of frequencies sampled within the band gap).
Above, the electric field $\tilde{\mathbf{E}}_{j}$ is obtained from the uniform current source $\tilde{\mathbf{J}}_{j}$ by inverting the equation 
\begin{equation}
\label{eq:Maxwellagain}
\mathcal{M}^{\mathbf{E}}(\epsilon,\mu,\omega)\tilde{\mathbf{E}}_{j}(\omega,\mathbf{r}) = i\omega\tilde{\mathbf{J}}_{j}(\mathbf{r}) = i\omega\frac{\hat{e}_j}{V_{pc}}
\end{equation}
where $V_{pc}$ is the volume of the primitive unit cell,
\begin{equation}
\label{eq:MaxwellopE}
\mathcal{M}^{\mathbf{E}}(\epsilon,\mu,\omega) = \mathbf{\nabla}\times\frac{1}{\mu(\mathbf{r})}\mathbf{\nabla}\times~ -~ \epsilon(\mathbf{r})\omega^2
\end{equation}
is the Maxwell operator for the electric field ($\epsilon(\mathbf{r})$, $\mu(\mathbf{r})$, and $\omega$ are the permittivity, permeability, and frequency, respectively), and $\hat{e}_j$ is a unit vector in the direction of the uniform current source. 

\begin{figure*}[ht!] 
\centering
\includegraphics[width=0.99\textwidth]{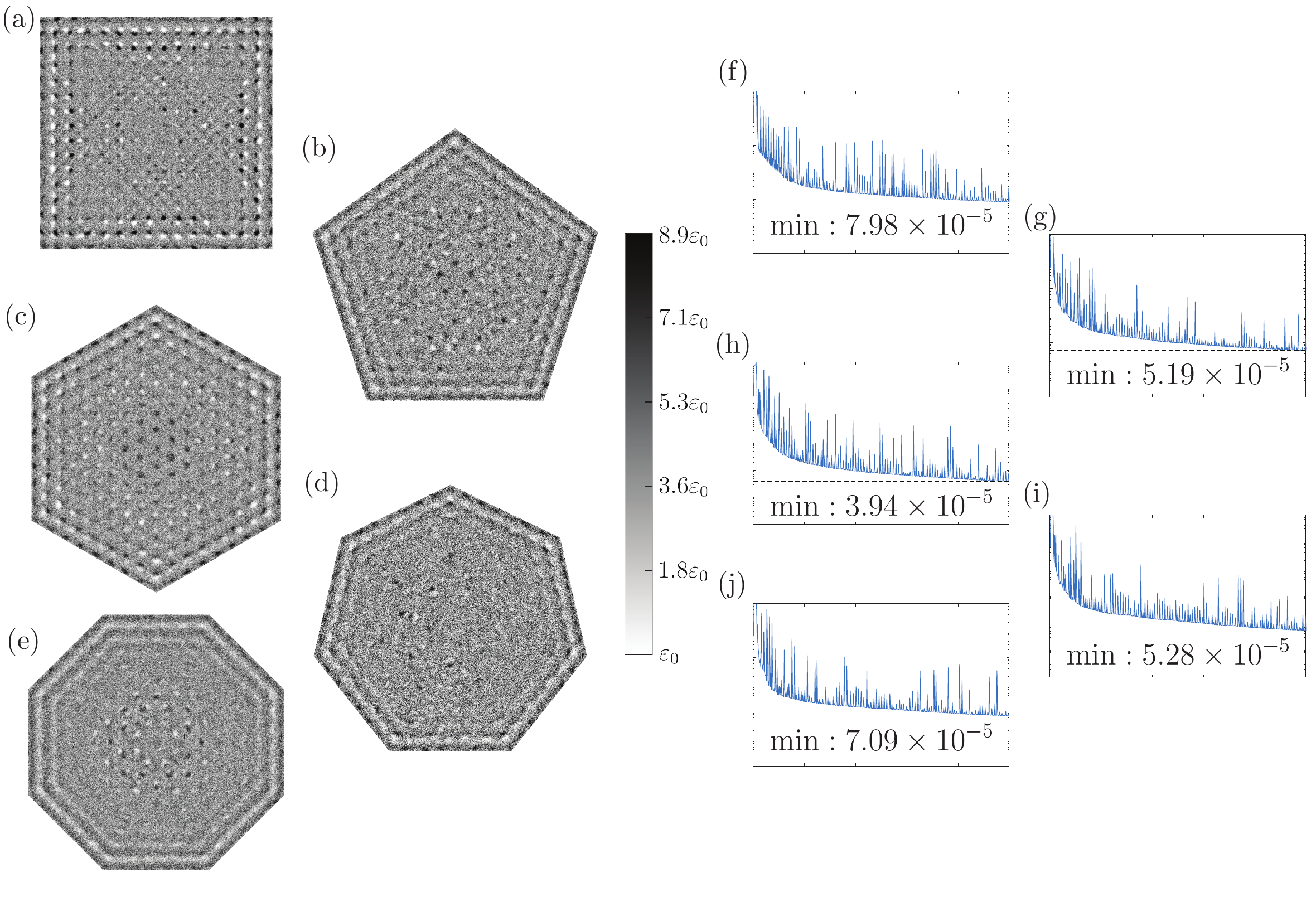}
\caption{Regular polygons of area $81a^2$ optimized for TM-polarized sources with a grid-point resolution (GPR) value of 100. We employed $\omega_0 = 0.8\cdot 2\pi c/a$ and $\Delta \omega = \omega_0/10$ in generating the designs for a square (a), a pentagon (b), a hexagon (c), a heptagon (d), and an octagon (e). The convergence of the photonic density of states figure of merit normalized by the corresponding quantities for vacuum is shown for the square in (f), the pentagon in (g), the hexagon in (h), the heptagon in (i), and the octagon in (j). The permittivity values were confined to the range $\frac{\epsilon(\mathbf{r})}{\epsilon_0} \in [1,8.9]$ and each grid point in a design was randomly initialized. The horizontal axis of the plots (f)-(j) shows the interval from 0 to 500 iterations and the vertical axis of the plots (f)-(j) the range from $10^{-6}$ to 1. The minimum value obtained by the normalized objective is indicated in (f)-(j) using a dashed line and with the numerical value.}
\label{fig:TM_regularshapes}
\end{figure*}

\begin{figure*}[ht!] 
\centering
\includegraphics[width=0.99\textwidth]{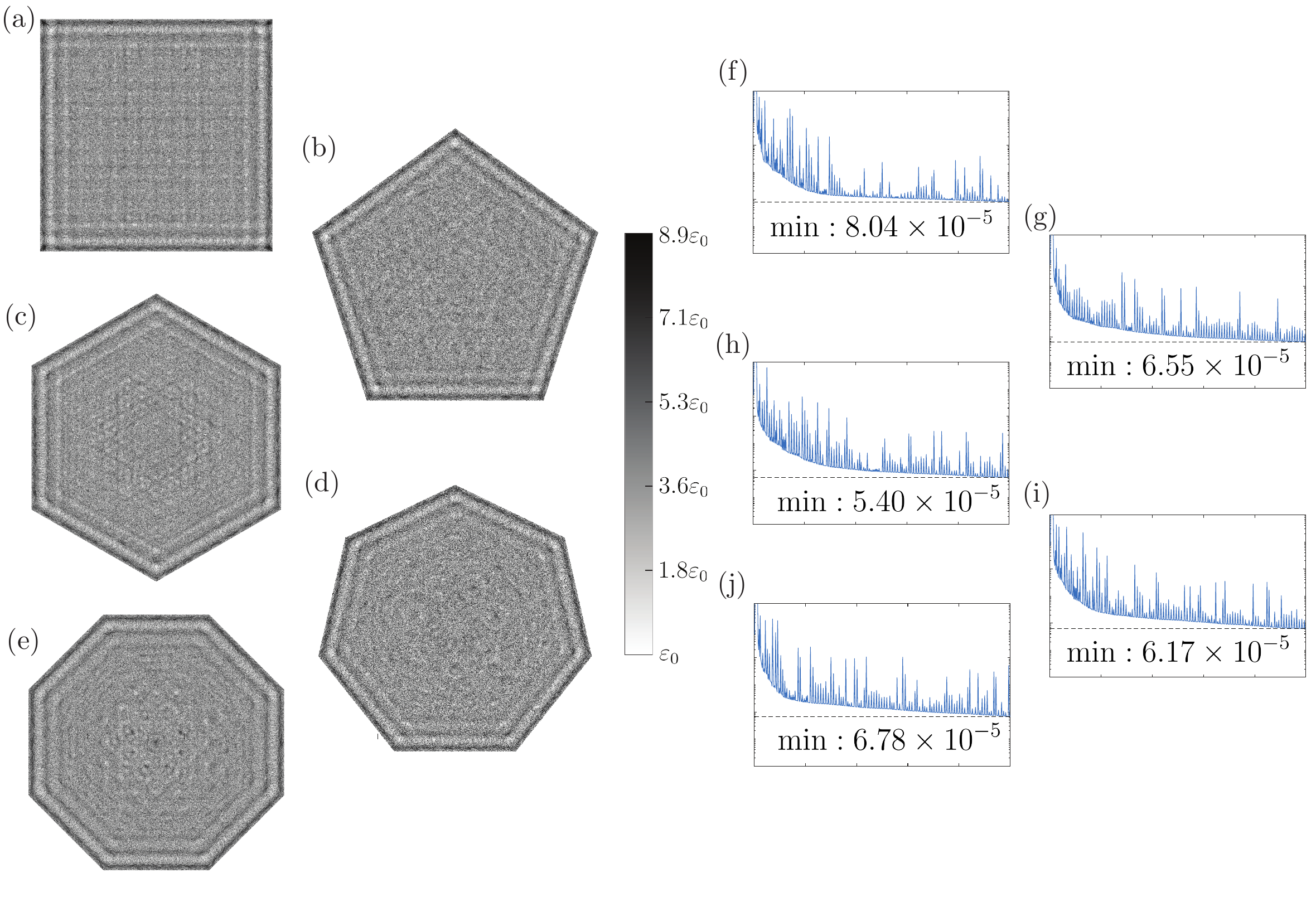}
\caption{Regular polygons of area $81a^2$ optimized for TE-polarized sources with a GPR value of 100. We employed $\omega_0 = 0.8\cdot 2\pi c/a$ and $\Delta \omega = \omega_0/10$ in generating the designs for a square (a), a pentagon (b), a hexagon (c), a heptagon (d), and an octagon (e). The convergence of the photonic density of states figure of merit normalized by the corresponding quantities for vacuum is shown for the square in (f), the pentagon in (g), the hexagon in (h), the heptagon in (i), and the octagon in (j). The permittivity values were confined to the range $\frac{\epsilon(\mathbf{r})}{\epsilon_0} \in [1,8.9]$ and each grid point in a design was randomly initialized. The horizontal axis of the plots (f)-(j) shows the interval from 0 to 500 iterations and the vertical axis of the plots (f)-(j) the range from $10^{-6}$ to 1. In (f)-(j), the minimum value obtained by the normalized objective is indicated both by a dashed line and with the numerical value.}
\label{fig:TE_regularshapes}
\end{figure*}

\begin{figure*}[ht!] 
\centering
\includegraphics[width=0.99\textwidth]{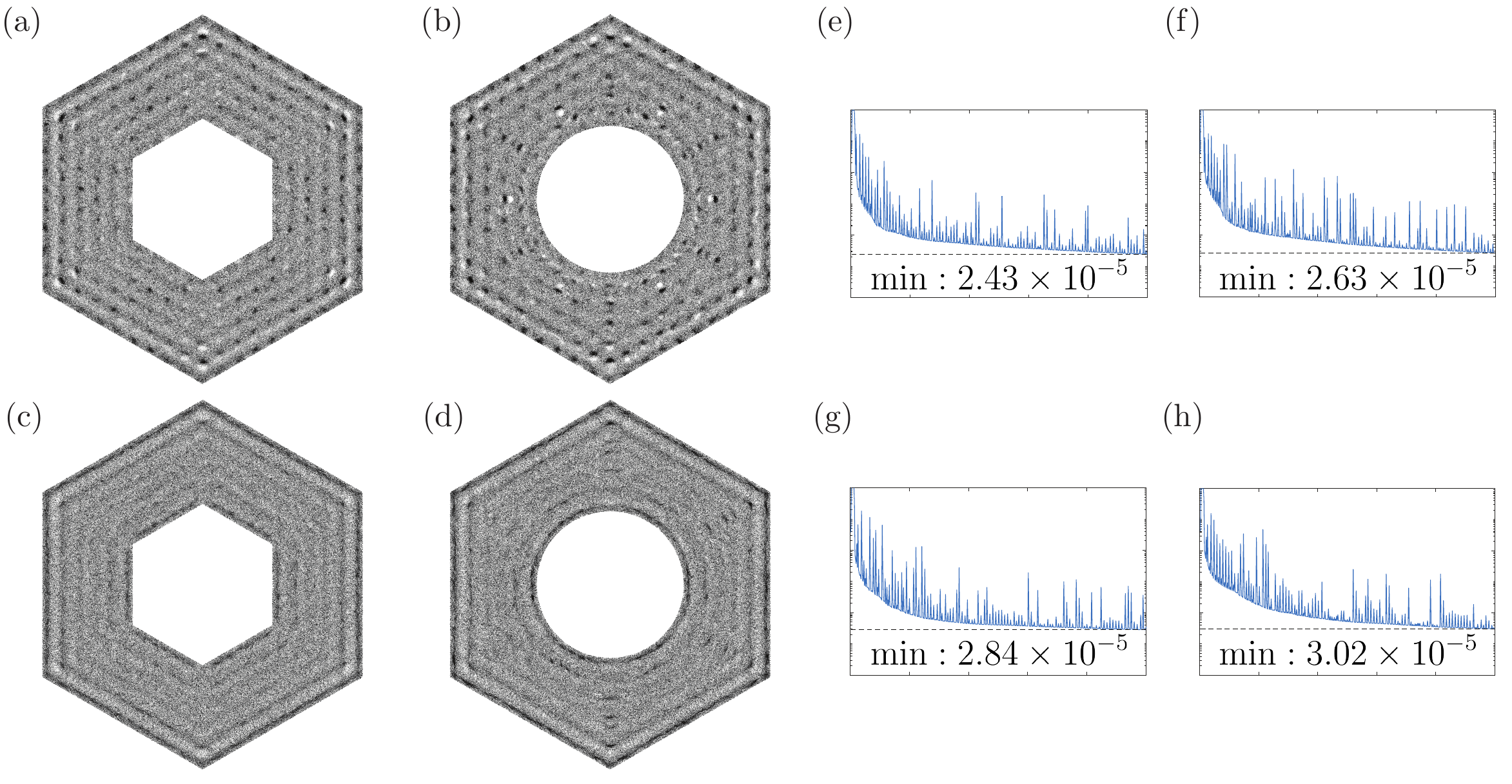}
\caption{Hexagons with cavities optimized for TM- and TE-polarized sources with a GPR value of 100. We employed $\omega_0 = 0.8\cdot 2\pi c/a$ and $\Delta \omega = \omega_0/10$ in generating the designs using a TM-polarized source for a hexagon with a hexagonal cavity (a) and a hexagon with a circular cavity (b), and using a TE-polarized source for a hexagon with a hexagonal cavity (c) and a hexagon with a circular cavity (d). The convergence of the photonic density of states figure of merit normalized by the corresponding quantity for vacuum is shown in (e) for (a), (f) for (b), (g) for (c), and (h) for (d). The permittivity values were confined to the range $\frac{\epsilon(\mathbf{r})}{\epsilon_0} \in [1,8.9]$ and each grid point in a design was randomly initialized (see the color bar in Fig. \ref{fig:TM_regularshapes}). The horizontal axis of the plots in (e)-(h) show the interval from 0 to 500 iterations and the vertical axis of the plots in (e)-(h) the range from $10^{-6}$ to 1. In (e)-(h), the minimum value obtained by the normalized objective is indicated both with a dashed line and with the numerical value. The annular regions all had area  $81a^2$.}
\label{fig:TM_TE_shapes}
\end{figure*}

We consider the objective in Eq. (\ref{eq:finalresidue}) in optimizing various regular polygons in Figs. \ref{fig:TM_regularshapes} and \ref{fig:TE_regularshapes}. The designs were randomly initialized with permittivity values in the range $\frac{\epsilon(\mathbf{r})}{\epsilon_0} \in [1,8.9]$, with a central frequency $\omega_0 = 0.8\cdot 2\pi c/a$, and a band gap $\Delta \omega = \omega_0/10$. A grid-point resolution (GPR) of 100 was used, which is the number of grid points used to resolve a length $a$. The quantity $c$ is the speed of light, $\epsilon_0$ is the permittivity of free space, and $a$ is a length scale that we can set to 1 due to the scale invariance of Maxwell's equations~\cite{joannopoulos_photonic_2011}. We modify ceviche code~\cite{hughes_forward-mode_2019} to carry out our optimizations with the NLopt package~\cite{johnson_nlopt_2007} employing the Method of Moving Asymptotes (MMA) algorithm~\cite{svanberg_class_2002}. To allow for comparisons between polygons of different numbers of sides, the areas of the polygonal design regions were fixed at $81a^2$ and were carved out of a $12a\times12a$ design region. The choice of $81a^2$ for the area allows our designs to exist within the regime where the length of a linear dimension of a square polygonal design region is greater than the corresponding Bragg length~\cite{bahulikar_structure_2025,joshi_structure_2025}. We see in Figs. \ref{fig:TM_regularshapes} and \ref{fig:TE_regularshapes} that the objective attains a minimum for structures with 6-fold symmetry, consistent with known results~\cite{joannopoulos_photonic_2011}. Specifically, since a hexagonal unit cell has higher rotational symmetry in 2D compared to a square unit cell, the corresponding Brillouin zone is more circular in shape so that the gap can have a large overlap in different directions (ie. when evaluated at different wave vectors)~\cite{joannopoulos_photonic_2011}. We also investigated hexagons with hexagonal or circular cavities (shown in Fig. \ref{fig:TM_TE_shapes}) and found that suppression of the photonic density of states was improved compared to the structures without a cavity. The structures with a hexagonal cavity performed better than the structures with a circular cavity, which may find application in the design of fiber-optic cables. To ensure a fair comparison, the outer hexagon in Fig. \ref{fig:TM_TE_shapes} was assigned an area of $100a^2$ and the cavity was assigned an area of $19a^2$ so that the annular structure obtained by combining the outer hexagon with the cavity would have an area of $81a^2$.

\begin{figure*}[ht!] 
\centering
\includegraphics[width=0.99\textwidth]{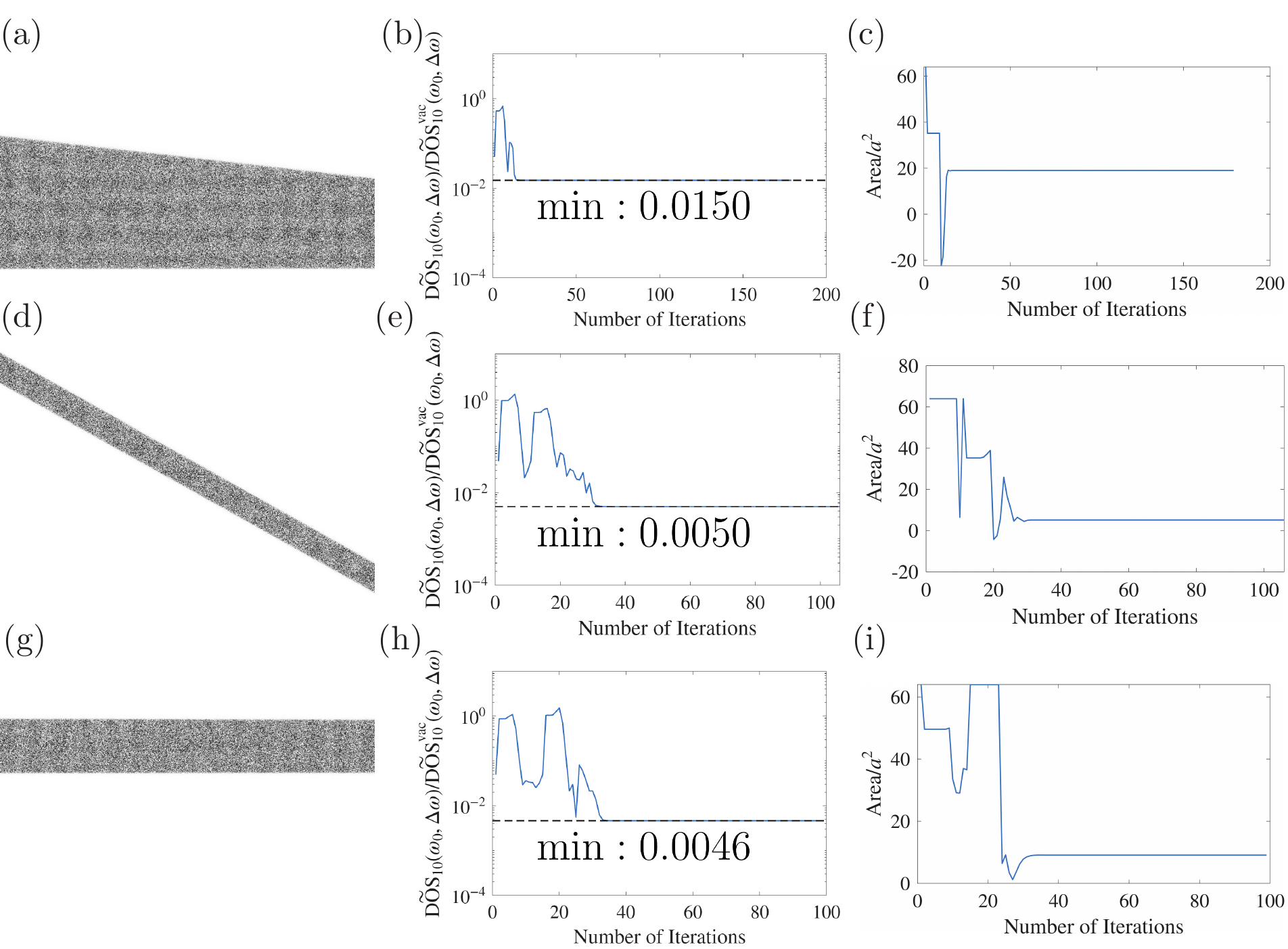}
\caption{Optimization of designs with two control points for the upper boundary function and two for the lower boundary function using the continuous design mask proposed in Eq. (\ref{eq:design_mask}) with $\alpha(y) = \arctan(y)/\pi$. We employed a GPR of 100, $\omega_0 = 0.8\cdot 2\pi c/a$, and $\Delta \omega = \omega_0/10$ in optimizing the designs in an $8a\times8a$ design region. For $\kappa = 70$, the design, the convergence of the figure of merit, and the progression of the area are shown in (a), (b), and (c), respectively. The corresponding quantities for $\kappa = 80$ and $\kappa = 90$ are shown in (d), (e), and (f), and (g), (h), and (i), respectively. TM polarization was used. See the color bar in Fig. \ref{fig:TM_regularshapes} for the permittivity values. The optimizations terminated with a criteria of $10^{-8}$ for the relative tolerance of the optimization parameters~\cite{johnson_nlopt_2007}. A dashed line indicates the minimum of the normalized objective function and the numerical value is also displayed in (b), (e), and (h).}
\label{fig:irregular_designs}
\end{figure*}

We next proceed to optimize $y$-simple geometries by applying a continuous design mask instead of a discrete one. For a $y$-simple geometry, this design mask has the form 
\begin{equation}
\gamma(x,y) = \alpha\left(\kappa\left(f(x)-y\right)\right)+\alpha\left(\kappa\left(y-g(x)\right)\right),
\label{eq:design_mask}
\end{equation}
where we employ $\alpha(y) = \arctan(y)/\pi$ (generally $\alpha$ can be any sigmoid function such that $\alpha(y) \in [-1/2,1/2]$), $\kappa$ is a constant, and the functions $f(x)$ and $g(x)$ describe the upper bounds and lower bounds, respectively, of the design region as a function of $x$. For an $x$-simple geometry, one would interchange $x$ and $y$ in Eq. (\ref{eq:design_mask}). In performing our optimizations, we first considered two control points~\cite{hughes_isogeometric_2005,seo_isogeometric_2010,seo_shape_2010,qian_isogeometric_2011} for the upper boundary function $f(x)$ and two for the lower boundary function $g(x)$ placed at the extremal values of $x$ with linear interpolation (B-splines of degree 1~\cite{qian_isogeometric_2011}) between the control points. In setting up our optimizations, we randomly initialized entire $8a\times8a$ design regions with permittivity values in the range $\frac{\epsilon(\mathbf{r})}{\epsilon_0} \in [1,8.9]$. There were no restrictions on the area of the final design region. When $\kappa = 80$, we see in Fig. \ref{fig:irregular_designs} that the design region transforms to a supercell of a hexagonal unit cell to within 1$^\circ$ as expected from the literature~\cite{joannopoulos_photonic_2011}. Indeed, we observe in Fig. \ref{fig:irregular_designs} a transition around $\kappa = 80$ where suppression of the photonic density of states figure of merit becomes more pronounced, which is consistent with our choice of $\omega_0 = 0.8\cdot 2\pi c/a$ and GPR = 100. For $\kappa = 70$, the design region transforms to a more general quadrilateral, while for $\kappa = 90$ the design region transforms to a rectangular supercell within the original square design region. We note that suppression of the photonic density of states improves slightly for $\kappa = 90$ compared to the suppression for $\kappa = 80$. This result is likely due to the fact that a hexagonal supercell admits a smaller area within the design region compared to a rectangular supercell (a supercell of a square primitive unit cell), which therefore provides more freedom for optimization within the supercell. We have observed greater suppression for other values of $\kappa$ such as $\kappa = 120$, but the resulting areas were unphysical (corresponding to less than one grid point along at least one dimension).

\begin{figure*}[ht!] 
\centering
\includegraphics[width=0.99\textwidth]{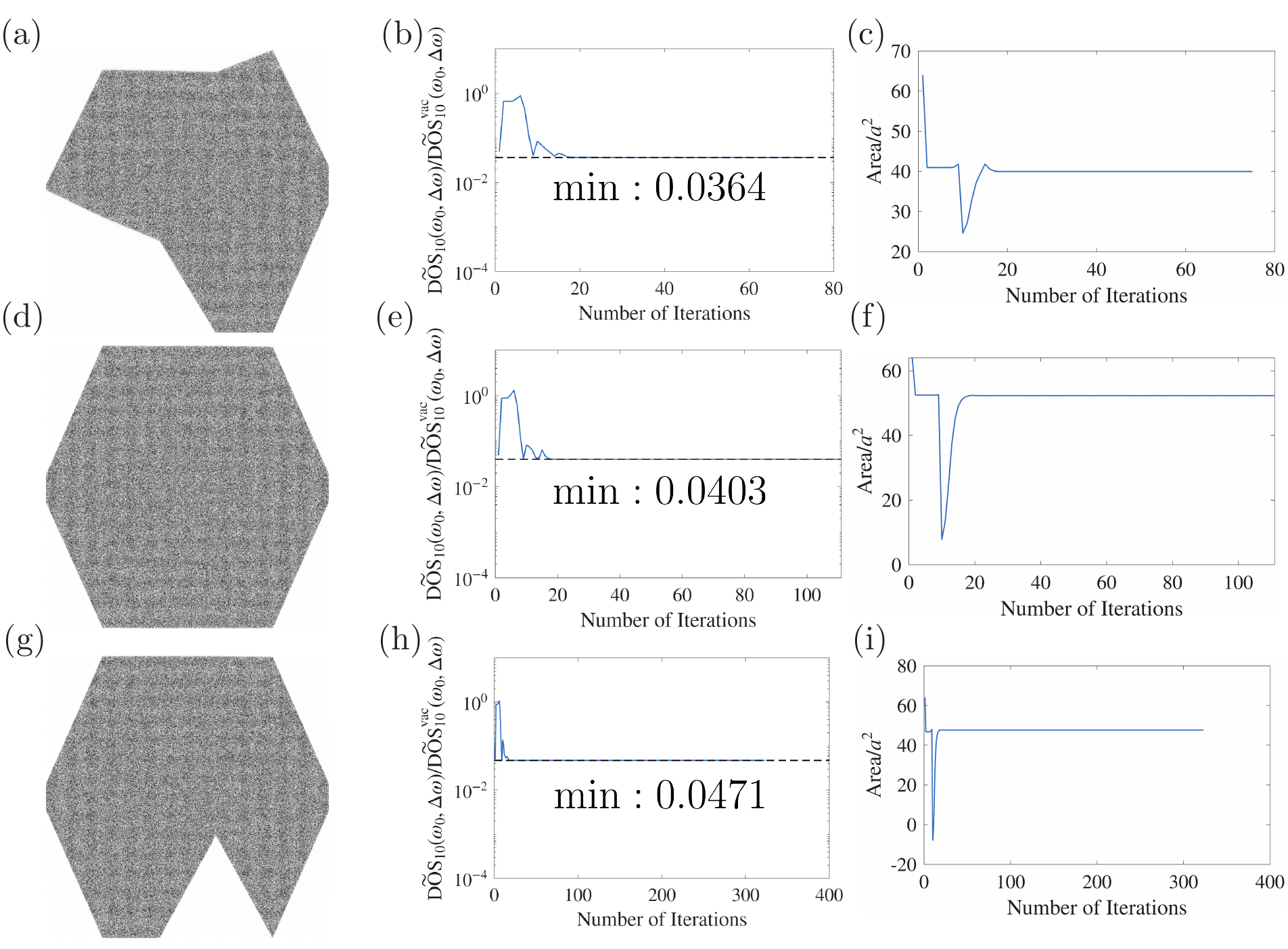}
\caption{Optimization of designs with six control points for the upper boundary function and six for the lower boundary function using the continuous design mask proposed in Eq. (\ref{eq:design_mask}) with $\alpha(y) = \arctan(y)/\pi$. We employed a GPR of 100, $\omega_0 = 0.8\cdot 2\pi c/a$, and $\Delta \omega = \omega_0/10$ in optimizing the designs in an $8a\times8a$ design region. As above, for $\kappa = 70$, the design, the convergence of the figure of merit, and the progression of the area are shown in (a), (b), and (c), respectively. The corresponding quantities for $\kappa = 80$ and $\kappa = 90$ are shown in (d), (e), and (f), and (g), (h), and (i), respectively. TM polarization was used. See the color bar in Fig. \ref{fig:TM_regularshapes} for the permittivity values. The optimizations terminated with a criteria of $10^{-8}$ for the relative tolerance of the optimization parameters~\cite{johnson_nlopt_2007}. The minimum value of the normalized optimization objective is indicated in (b), (e), and (h) with a dashed line and with the numerical value.}
\label{fig:irregular_designs_Nctrl11}
\end{figure*}

We have also investigated increasing the number of control points associated with the boundaries from two to six so that the region would be divided evenly along $x$ into five $y$-simple subregions, as shown in Fig. \ref{fig:irregular_designs_Nctrl11}. We recover structures that resemble hexagons for $\kappa = 80$ and $\kappa = 90$ and we find that the structure (g) in Fig. \ref{fig:irregular_designs_Nctrl11} appears to be attempting to arrange itself in such a manner as to create a cavity, consistent with the enhanced suppression in Fig. \ref{fig:TM_TE_shapes} compared with Figs. \ref{fig:TM_regularshapes} and \ref{fig:TE_regularshapes}. Such cavity-inducing arrangements are, however, inhibited by the $y$-simple nature of the optimization regions.

Our results ultimately provide a mechanism for discovering photonic crystals of arbitrary symmetry, appropriate to a given set of material parameters, without the exhaustive searches that would become daunting in three dimensions. In order to accomplish this objective, we show that topology optimization can be performed simultaneously using continuous relaxations of the permittivity and a formalism analogous to level-set approaches. By transforming a discretely defined design mask into a continuous quantity we show rapid convergence of our objective to produce designs that recover known crystal symmetries. We also find that hexagonal structures with hexagonal cavities perform better than hexagonal structures with circular cavities, a result that may find application in the design of fiber-optic cables. 

\section{Acknowledgements}
R.K.D. acknowledges financial support that made this work possible from the College of Engineering and Computer Science of Syracuse University. The authors also acknowledge that the work reported on in this paper was substantially performed using Zest, the Syracuse University research computing high-performance computing cluster. Finally, we wish to acknowledge fruitful conversations with Alejandro W. Rodriguez.

 % Display the acknowledgments section

% Bibliography
% *****************************************************

\bibliography{refs_PC}
\end{document}